\documentclass[]{jfm}
\usepackage{graphicx, siunitx, csquotes, booktabs}
\usepackage{newtxtext,newtxmath,natbib, hyperref}
\newcommand{\RomanNumeralCaps}[1]
\linenumbers

\shorttitle{A drift velocity mediated eddy diffusivity model}
\shortauthor{O. B. Shende, L. Storan, and A. Mani}

\title{A model for drift velocity mediated scalar eddy diffusivity in homogeneous turbulent flows}

\author{
Omkar B. Shende\aff{1} \corresp{\email{oshende@stanford.edu}},
Liam Storan\aff{2}
\and Ali Mani\aff{1}
}

\affiliation{
\aff{1}Department of Mechanical Engineering, Stanford University,
Stanford, CA 94305, USA
\aff{2}Department of Applied Physics, Stanford University, Stanford, CA 94305, USA
}

\begin{document}
\maketitle
\begin{abstract} 

\noindent Low Stokes number particles at dilute concentrations in turbulent flows can reasonably be approximated as passive scalars. The added presence of a drift velocity due to buoyancy or gravity when considering the transport of such passive scalars can reduce the turbulent dispersion of the scalar via a diminution of the eddy diffusivity. In this work, we propose a model to describe this decay and use a recently developed technique to accurately and efficiently measure the eddy diffusivity using Eulerian fields and quantities. We then show a correspondence between this method and standard Lagrangian definitions of diffusivity and collect data across a range of drift velocities and Reynolds numbers. The proposed model agrees with data from these direct numerical simulations, offers some improvement to previous models in describing other computational and experimental data, and satisfies theoretical constraints that are independent of Reynolds number. 

\end{abstract}

\section{Introduction} 
The motion of particles in a turbulent carrier fluid subject to exogenous body forces manifests in myriad applications such as ash settling \citep{mingotti_2020}; energy production \citep{ishii_1977, guet_2006}; and bubbly wakes \citep{carrica_1999}. These particles may experience multiphysics including nucleation, coalescence, dissolution, and growth, but their dispersion by energetic eddies of the background flow naturally lends itself to the field of turbulence modelling. In the framework of population balance equations, as reviewed in \citet{shiea_2020}, particles can be segregated into size classes associated with a \emph{drift velocity}, $u_d$, relative to the background flow with root-mean-square velocity $u_{rms}$. In nature, $u_d/u_{rms}$ can be large; a Hinze-scale \qty[per-mode=symbol]{1}{\mm} bubble may rise at \qty[per-mode=symbol]{12}{\cm \per \s}, while a characteristic upper ocean turbulence velocity is closer to \qty[per-mode=symbol]{2}{\cm \per \s}. \citep{detsch_1989, darso_2014}

For dilute, low Stokes number flows with negligible particle-particle interactions, the particle concentration -- or void concentration -- fields can be described with the passive scalar advection-diffusion equation. \citep{moraga_2003} Each class can be evolved separately and forces like buoyancy and gravity are accounted for with additional velocity components from relations like the drag law of \citet{schiller_1935}. For sedimenting particles, this drift is aligned with the gravity vector; in the case of bubbly flows, it is generally anti-parallel. 

Solving for full-resolution scalar evolution when only a mean state is required to determine quantities of engineering interest is prohibitively expensive. Averaging to find the mean occurs in homogeneous spatio-temporal dimensions, defined through Reynolds-averaging or filtering in the large-eddy simulation context. When averaging is applied to the Navier-Stokes and advection-diffusion equations that govern the evolution of scalar-laden fluid flows, the turbulent scalar flux, given by $\overline{u_i c}$, appears. Here, $u_i$ and $c$ are the fluctuating velocity and scalar fields, respectively, and $\overline{\bullet}$ is an average. This term represents unresolved scalar transport by turbulence and its closure is essential to making transport simulations tractable.

A common model form for the turbulent scalar flux is gradient diffusion, written analogously to Fickian diffusion as $ \overline{u_i c} = - D_{ij} \frac{\partial \overline{C}}{\partial x_j}$, with $D_{ij}$ representing eddy diffusivity and $C$ the full scalar field. Foundational work has shown eddy diffusivity decays with increased drift, but few extant models for capturing the flux are algebraic closed-form expressions. \citep{yudine_1959, moraga_2003, reeks_2021} An exception is \citet{csanady_1963}, which derives an expression for particle diffusivity scaling as a function of $u_d$ from theoretical arguments about the form and relevant parameters of the velocity autocorrelation, but \citet{squires_eaton_1991} showed disparities between it and measured experimental and computational turbulent data.

As \citep{csanady_1963} serves as a starting point for more complex models (\textit{e.g.} \citep{wang_1993}) and particle-laden turbulence is still a ripe topic (\textit{e.g.} \citep{berk_coletti_2021}), we wish to revisit this problem using the recently developed Macroscopic Forcing Method (MFM) to calculate eddy diffusivity. \citep{mani_2021, shirian_2022} In general, the eddy diffusivity is a non-local spatial and temporal operator, but when there is separation of scales between the large-eddy size and the scalar cloud size, as is relevant for this problem, measurement of a single local coefficient of eddy diffusivity, denoted $D^0$ and sometimes called a dispersion coefficient, suffices. \citet{mani_2021} showed that $D^0$ is the leading-order moment of the full eddy diffusivity operator and \cite{shende_2022} used it to predict scalar transport. Furthermore, it is the first term in the Kramers-Moyal approximation of the integro-differential kernel that defines the true eddy diffusivity operator.

In this work, we propose a model using measurements of eddy diffusivity by MFM in numerical simulations of scalar fields driven by homogeneous, isotropic turbulence (HIT) over a range of drift velocities. MFM has more favorable computational costs when compared with a similar Lagrangian method for this \emph{a priori} modelling approach. The proposed model better fits empirical data and theoretical constaints for tracers subject to drift. 

\section{Model problem}  

Consider a triply-periodic cubic domain of HIT. When $u_d = 0$, the diffusivities along the three principal axes -- $D^0_{11}$, $D^0_{22}$, and $D^0_{33}$ -- are all equal to $D^0$. If we impose a non-negative drift velocity in the $x_1$ direction, the symmetry of the setup is broken such that only the diffusivities in the $x_2$ and $x_3$ directions are equal and all are affected by drift. While the derivation herein reaches conclusions similar to others (\textit{e.g.} \citep{csanady_1963, squires_eaton_1991, mazzitelli_2004}), we distinctly adopt an Eulerian perspective.

For $u_d = 0$, we begin with the Lagrangian formulation of eddy diffusivity of \citet{taylor_1921}, with the scalar represented by tracer particles with position $X_i(t)$ and velocity $V_i(t)$. \citet{taylor_1921} writes the eddy diffusivity in the $x_1$ direction of a ensemble of such particles, in the long time limit. If the velocity field is statistically stationary, this can be expressed as

\begin{equation}  \label{eqn:lagrangianD0} 
   D^0_{11} = \int^\infty_0 \langle  (V_1(\tau+t)~V_1(t)\rangle~d\tau.
\end{equation}

\noindent where $\langle \bullet \rangle$ represents an average over the ensemble. In the high P\'eclet number ($Pe$) limit, where molecular diffusivity is far smaller than eddy diffusivity, $V_i(t)$ is the flow velocity at a given particle's position. Therefore, we can express the underlying Eulerian velocity field component, $u_1$, as a function of the full three-dimensional tracer position, $\mathbf{X}$. This yields

\begin{equation}    \label{eqn:diffusionintegral}
    D^0_{11} =  \int^\infty_0 \langle u_1(\mathbf{X}(t),t) ~u_1(\mathbf{X}(t+\tau),t+\tau) \rangle~d\tau.
\end{equation}

We define a standard complementary characteristic turbulence length scale as

\begin{equation}    \label{eqn:lengthintegral}
    L_{11} = \frac{1}{u^2_{rms}} \int^\infty_0 \langle u_1(x_1, x_2, x_3,t)~u_1(x_1+r,x_2,x_3,t) \rangle~dr,
\end{equation}

\noindent where $r$ is a spatial offset, $\langle \bullet \rangle$ now represents an average over all independent variables, and $\tau$ is a temporal offset. Here, $u_{rms}$ normalizes the velocity autocorrelation at zero displacement. 

Classical Brownian motion analysis posits that scalar diffusivity at infinite time scales as $u_{rms}^2 \tau_k$, where $\tau_k \equiv L_k/u_{rms}$ denotes some timescale of the underlying flow development and $L_k = 2L_{11}$ is the large-eddy length scale. Consider the scalar field now with some constant drift velocity, $u_d$, with respect to the flow: for very small $u_d$, the fundamental turbulence statistics felt by a scalar parcel are not largely affected and the classical model holds. In the opposite limit that $u_{rms} << u_d$ such that $L_{11}/u_d << \tau_k$, however, this picture is not appropriate.

In this limit, the scalar field transits the turbulence field at a very fast timescale $\tau_d = L_{11}/u_d$, such that scalar particles drift before the local flow has evolved. The turbulence can therefore be considered frozen compared to the evolution of the scalar field for computing equation \ref{eqn:diffusionintegral}, and a leading-order approximation for the differential translation of a scalar parcel is $\Delta X_1 = dr \approx u_d~dt$. Thus, the addition of a very large drift is equivalent to examining a translating inertial coordinate frame with respect to the frozen flow, akin to Taylor's hypothesis. These premises allow us, for very large $u_d$, to rewrite equation \ref{eqn:diffusionintegral} as:

\begin{equation} \label{eqn:derivationintegral}    
   \lim_{u_d \to \infty} D^0_{11} =  \int^\infty_0 \langle u_1(x_1, x_2, x_3,t)~u_1(x_1+u_d\tau,x_2,x_3,t) \rangle~d\tau. \\
\end{equation}

We have here posited that the change in particle position is purely due to drift, and over the time of $O(L_{11}/u_d) << \tau_k$ where the kernel is non-zero, the field $u_1$ does not change. If we now use a change of variable between the displacement and the drift velocity, we can write

\begin{equation}   
    =  \int^\infty_0 \langle u_1(x_1, x_2, x_3,t) ~ u_1(x_1+r,x_2,x_3,t) \rangle~d(r/u_d) =  u^2_{rms}~\frac{L_{11}}{u_d}. \\
\end{equation}

Here, $L_{11}$ is the correlation length scale in equation \ref{eqn:lengthintegral} and it requires the spatial correlation drop to zero in the domain. For the transverse diffusivities, $D^0_{22} = D^0_{33}$. Appealing to the isotropy of the underlying velocity fields, we can show that:


\begin{equation}   
    \lim_{u_d \to \infty} D^0_{22} =  \int^\infty_0 \langle u_1(x_1, x_2, x_3,t) ~ u_1(x_1,x_2+r,x_3,t) \rangle~d(r/u_d) =  u^2_{rms}~\frac{L_{22}}{u_d}.  \\
\end{equation}




This asymptotic limit jibes with intuition, as a particle with infinite $u_d$ samples zero mean velocity over every time horizon in homogeneous turbulence. This decay of $D^0_{ii}$ with increasing $u_d/u_{rms}$, the \enquote{crossing trajectories} effect of \citet{yudine_1959} and \citet{csanady_1963}, persists in relatively high Reynolds number ($Re$) bubble experiments. \citep{mathai_2018}

A simple model form that captures the asymptotic limits of infinite and zero drift is:

\begin{equation} \label{eqn:D0model}
    \frac{D^0_{ii}}{D^0} \approx \left( 1 + \left(\frac{u_d~D^0}{L_{ii}~u^2_{rms}}\right)^{\alpha_{ii}} \right)^{-1/\alpha_{ii}} = \left( 1 + \left(\frac{u_d}{u^*_{ii}}\right)^{\alpha_{ii}} \right)^{-1/\alpha_{ii}}. 
\end{equation}

$\alpha_{ii}$ is a free parameter and $u_{ii}^*$ is an Eulerian \enquote{diffusion velocity} that competes with the drift. In incompressible flows, $L_{22} = L_{33} = L_{11}/2$. \citep{csanady_1963} The eddy diffusivity at large drift velocity can be written as $D^0_{ii}/D^0 = u^*_{ii}/u_d$. In contrast, \cite{csanady_1963} proposed 

\begin{equation} \label{eqn:csanady}
        \frac{D^0_{ii}}{D^0} \approx \left( 1 + \left(\frac{\gamma_{ii}~\beta~ u_d}{u_{rms}}\right)^2 \right)^{-1/2},
\end{equation}

\noindent where $\beta$ is defined with Lagrangian and Eulerian statistics and $\gamma_{22} = 2\gamma_{11} = 2$.

\section{Numerical setup}  

Direct numerical simulation (DNS) of forced incompressible HIT in a triply-periodic box of edge length $L_{box}$ provides the data for this test. Solved with continuity are the incompressible Navier-Stokes momentum equations with a linear forcing term. They are written as:

\begin{equation}
    \frac{\partial{u_i}}{\partial{t}} + \frac{\partial{u_i u_j}}{\partial x_j} = - \frac{1}{\rho} \frac{\partial p}{\partial x_i} + \nu \frac{\partial^2 u_i}{\partial x_j \partial x_j} + A\widetilde{u_i},
\end{equation}

\noindent where $A$ is a controller that maintains the turbulent kinetic energy, $k \equiv u_i u_i/2$, at a prescribed level, $\rho$ is the fluid density, and $\nu$ is the kinematic viscosity. This approach is similar to \citet{bassenne_2016}, but with a high-pass filtered velocity described in spectral space as $\hat{\widetilde{u_i}} = G(\mathbf{k})\hat{u_i}$, where $\mathbf{k}$ is the wavenumber magnitude and

\begin{equation}
    G(\mathbf{k}) =
   \begin{cases}
   0 & \mathbf{k} \leq 2 \\
   \frac{1}{2} - \frac{1}{2}\cos(\pi (\mathbf{k}-2))  & 2 < \mathbf{k} \leq 3 \\
   1 & \mathbf{k} > 3
   \end{cases}     
   .
\end{equation}

\begin{figure}
    \centering
\includegraphics[width=0.85\textwidth]{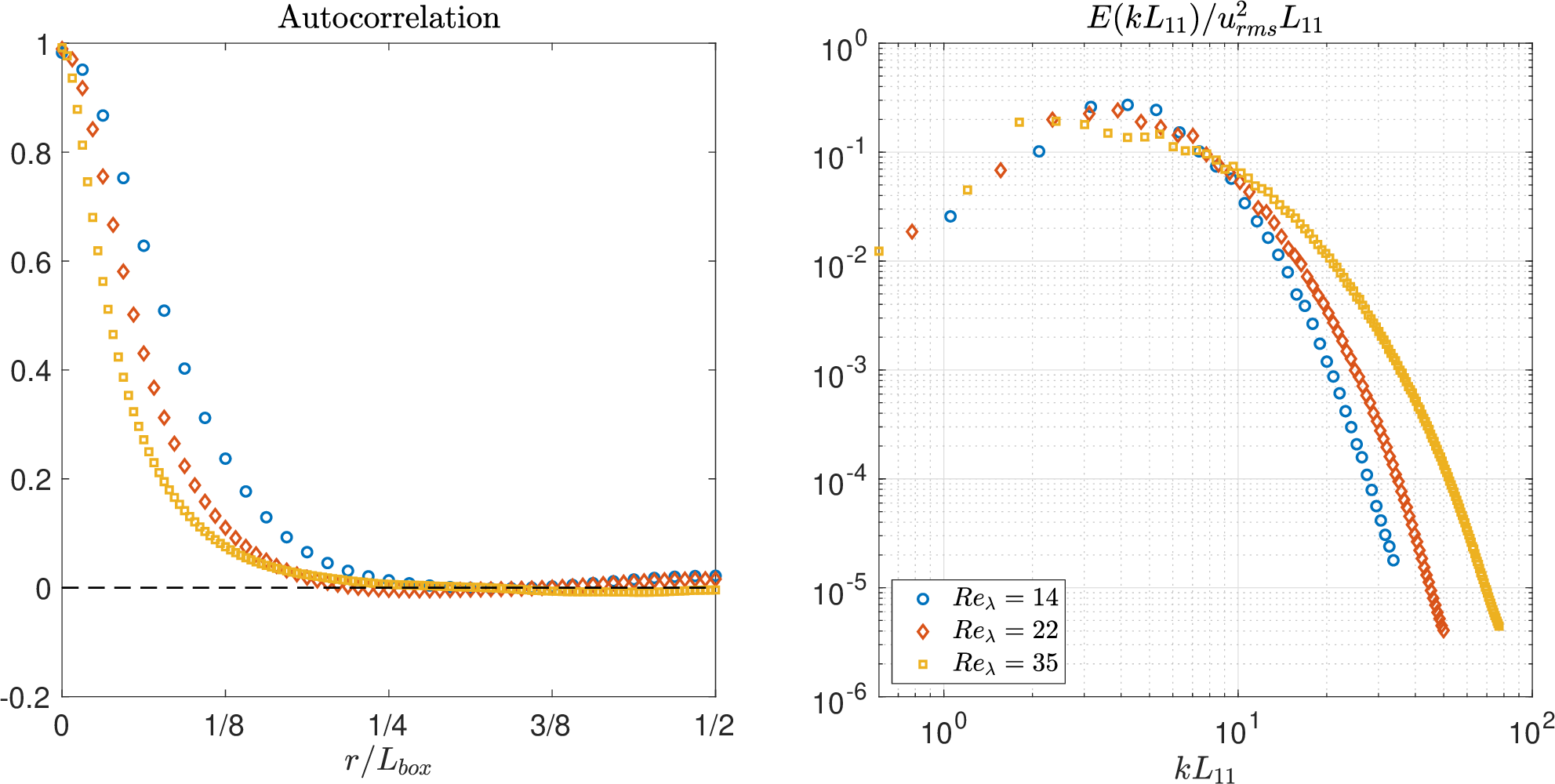}
    \caption{Instantaneous unnormalized velocity autocorrelations (L) and normalized energy spectra (R) for the cases in a $2\pi^3$ box, with filtered forcing preventing energy growth in the largest modes. Note that $u_{rms} = 1$ for all cases.}
    \label{fig:velcorr}
\end{figure}

In figure \ref{fig:velcorr}, plotted instantaneous energy spectra show decay at the largest scales for all considered $Re$. $L_k$, $L_{11}$, $L_{22}$, and $L_{33}$ are all less than $L_{box}/2$ by construction due to the filtered energy injection. As such, the simulations are not influenced by the cubic box shape or orientation, unlike the standard linear forcing method, as in \citet{rosales_2005}. 

To quantify eddy diffusivity, the MFM formulation of \citet{mani_2021} is used with code adapted from \citet{pouransari_2016} for solving the governing equations on a staggered $N^3$ grid with finite volume operators and a Runge-Kutta time-advancement scheme. MFM measures the response of the scalar field to an imposed macroscopic forcing by solving an additional equation for a scalar field with molecular diffusivity $D_m$. Following the procedure of \citet{shirian_2022} for finding $D^0$, we decompose the scalar field as $C = \overline{C} + c$ and add a macroscopic source term, $s$, to the scalar equation so that $c$ is governed by:

\begin{equation} \label{eqn:scalareqn}
    \frac{\partial{c}}{\partial{t}} + \nabla \bullet \left( ( \Vec{u} + \Vec{u_d} ) ~ c \right) = D_m \nabla^2 c -  (\Vec{u} + \Vec{u_d} ) \nabla \overline{C} + s(\Vec{x},t).
\end{equation}

Following the method of moments, the forcing maintains $\frac{\partial \overline{C}}{\partial x_i} = 1$. \citep{mani_2021} Setting $i = 1$ allows measurement of the \emph{axial} diffusivity in the direction of drift and $i = 2, 3$ allows quantification of \emph{transverse} diffusivity in directions perpendicular to drift. Table \ref{tab:D0} summarizes parameters swept to measure eddy diffusivity in the $x_1$ and $x_2$ directions. 

In all cases, $\nu=D_m$ and $u_{rms} = \sqrt{2k/3} = 1$. The underlying field being HIT means the root-mean-square value for each of the three velocity components is this $u_{rms}$. Once the velocity and scalar fields are fully developed, $D^0_{ii} = -\overline{u_ic}$ is post-processed from the turbulent scalar flux for a time of $O(200-500)~\tau_k$. Confidence intervals are calculated using the standard error of each mean statistic by constructing decorrelated samples of appropriate length compared to $\tau_k$. \citep{shirian_2023} For each simulation, $\Delta/\eta = \Delta/\lambda_B \approx 2$, where $\Delta$ is the grid spacing and $\eta \equiv \nu^{3/4}\epsilon^{-1/4}$ is the Kolmogorov length scale, $\lambda_B$ is the Batchelor scale, and $\epsilon$ is the energy dissipation rate. We also report $Re_\lambda \equiv \sqrt{\frac{15 u^4_{rms}}{\epsilon \nu}}$. 

\subsection{Comparison of MFM to Lagrangian formulae}  

We show equivalence in the MFM and Lagrangian definitions of eddy diffusivity. MFM measures $D^0$ by calculating the correlation of simulated scalar and velocity fields. The Eulerian-Lagrangian method (ELM), in contrast, simulates scalar particles in a background flow to calculate equation \ref{eqn:lagrangianD0}. Following \citet{falkovich_2001}, the particles are governed by $d\mathbf{X} = \mathbf{V}(t)~dt + \sqrt{2D_m}\mathbf{W}(t)~\sqrt{dt}$ where $\mathbf{W}$ is a zero-mean, unity-variance Wiener process.

\begin{figure}
    \centering
    \includegraphics[width=0.45\textwidth]{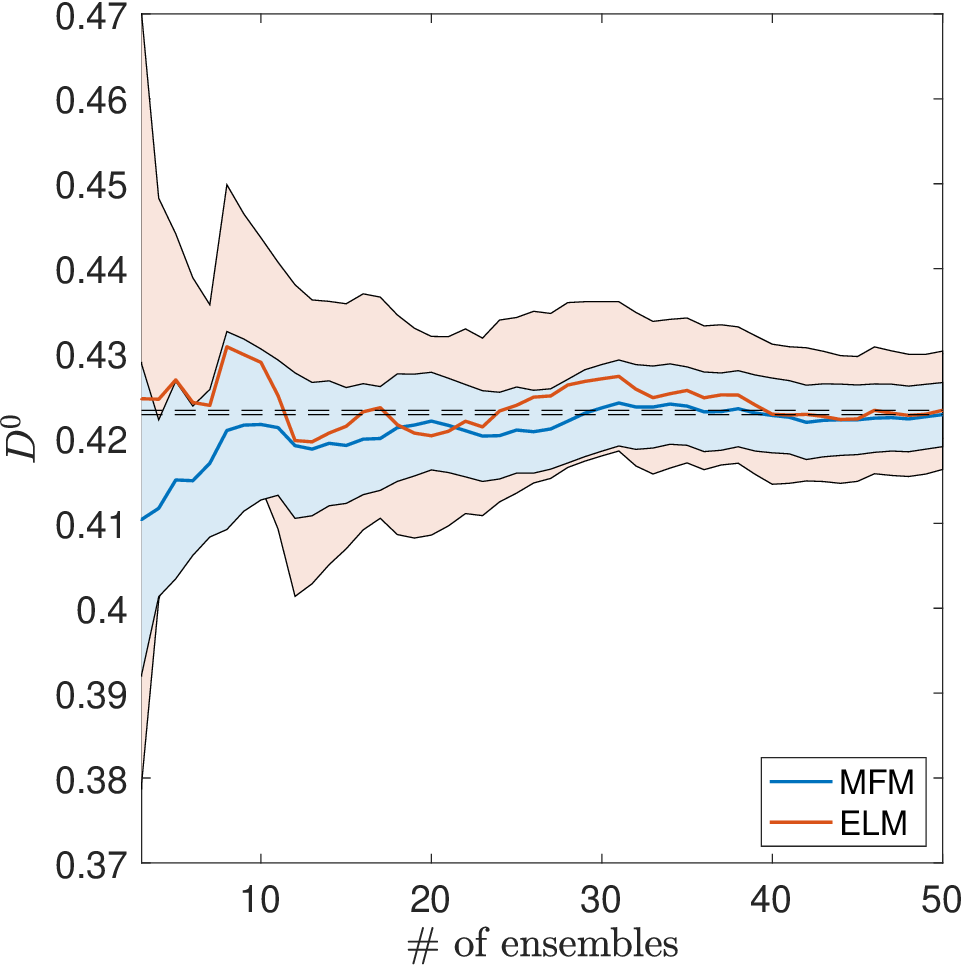}
    \caption{Mean estimates of $D^0$ for the $Re_\lambda =14.4$ case using MFM and ELM with 95\% confidence intervals showing convergence to the final estimate from each method (- -).}
    \label{fig:mfmelm}
\end{figure}

To assess this equivalence, both methods measure eddy diffusivity from the same HIT case of $L_{box} = 2\pi^3$ and $Re_\lambda = 14.4$ from table \ref{tab:D0}. In addition to the Eulerian DNS, MFM requires solution of equation \ref{eqn:scalareqn} on $O(10^5)$ mesh points. ELM requires the same flow-field DNS, along with Lagrangian trajectory simulations for $O(10^6)$ particles. Interpolation of velocity to particle locations uses modified Akima piecewise cubic Hermite functions. 

For the candidate turbulent flow, figure \ref{fig:mfmelm} compares the measured estimate of the mean value of the eddy diffusivity as a function of the number of ensembles considered. Each independent ensemble is of length $\approx 10 \tau_k$, collected from a fully developed DNS. The mean estimate from both methods is approximately the same for this realization, but the confidence intervals are quite dissimilar in their size. MFM provides more confident mean estimates for this application and is more scalable for simulations, as Eulerian fields are easier to distribute and solve with parallel computing than Lagrangian particles. An additional advantage of MFM not utilized here is that  MFM can find higher nonlocal moments of the diffusivity kernel beyond the local-limit leading-order moment. \citep{mani_2021} 

\section{Results and Discussion}  

\begin{table}
\centering
\begin{tabular}{c|ccccc}
$L_{box}, N^3$   & $2\pi, 64^3$ & $2\pi, 128^3$ & $2\pi, 256^3$  & $4\pi, 128^3$ & $8\pi, 256^3$ \\
$Re_\lambda$& 14.4  & 21.9  & 35.1 & 14.4  & 14.4 \\
$u^*_{11}/u_{rms}$  & 1.36  & 1.23  & 1.13  & 1.36 & 1.36  \\
$u^*_{22}/u_{rms}$  & 0.67  & 0.62  & 0.57  & 0.67 &  0.67 \\
\end{tabular}%
\caption{Summary of measured values relevant for the computation of $D^0_{11}$ and $D^0_{22}$.}
\label{tab:D0}
\end{table}

Equation \ref{eqn:D0model} is fitted to the mean diffusivity data from the cases of table \ref{tab:D0} as a function of $u_d$. An iterative method is used to determine $\alpha_{11}$ and $\alpha_{22}$ and the forthcoming section will show that the value of $\alpha_{ii}$ is largely invariant to the tested Reynolds numbers.

\begin{figure}
    \centering
    \includegraphics[width=0.9\textwidth]{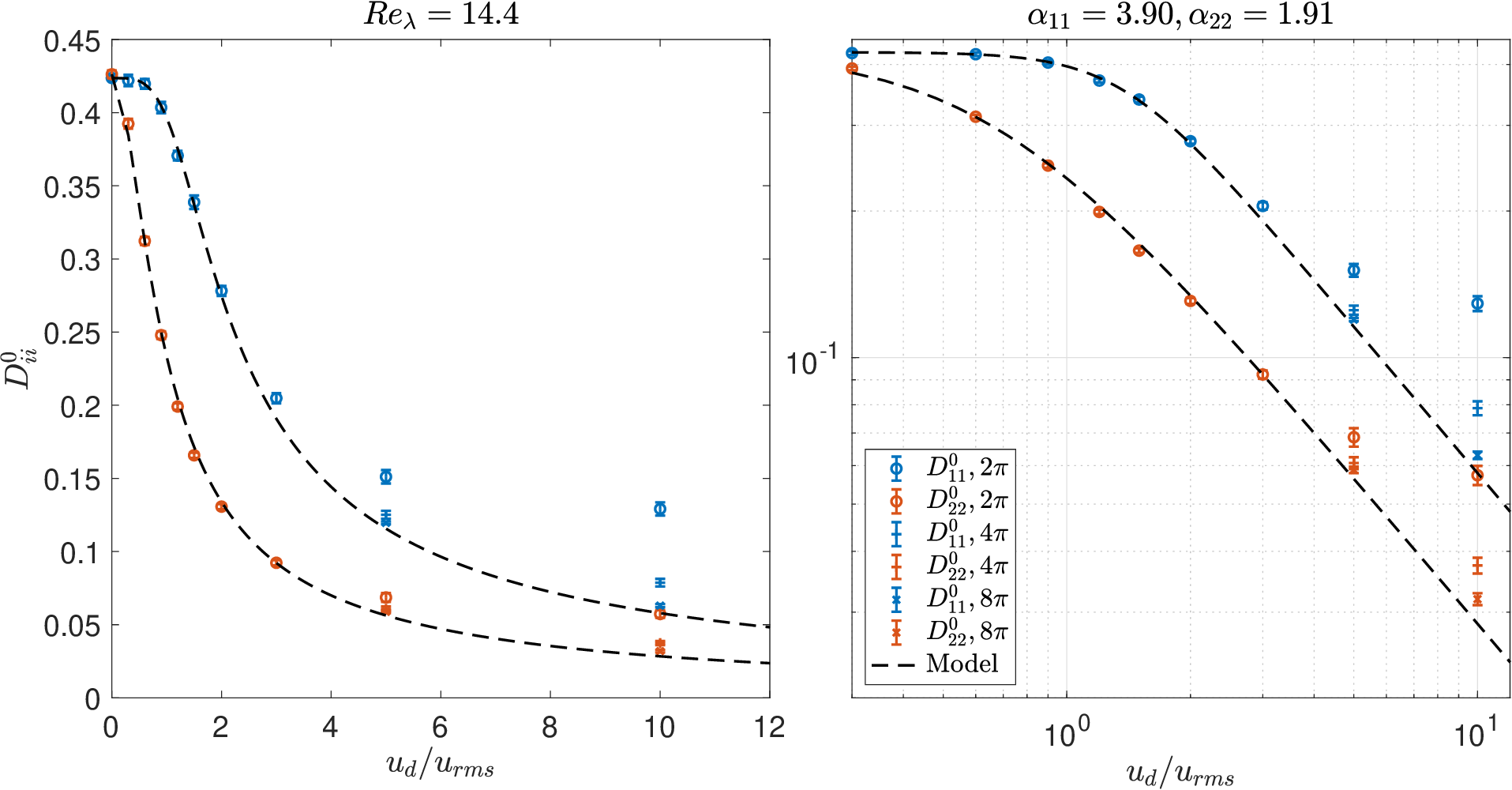}
    \caption{$D^0_{11}$ and $D^0_{22}$ for variable box sizes with $\alpha_{ii}$ indicated and model predictions. Data with 95\% confidence intervals in (L) standard and (R) log-log axes.}
    \label{fig:D0box}
\end{figure}

In figure \ref{fig:D0box}, $D^0_{11}$ and $D^0_{22}$ values are plotted for the $Re_\lambda = 14.4$ cases as a function of drift. The eddy diffusivity decay trends for the two directions differ for three values of $L_{box}$. At low drift velocities, the model accurately describes the data; however, at high drift, the $2\pi^3$ box data diverges from the analytic scaling. For a fixed box size, the computational value of $D^0_{ii}$ asymptotes to a nonzero value in the limit of large drift velocity, in violation of the analytical derivation presented earlier. This effect can be explained: in a finite periodic domain, a large drift velocity subjects a scalar particle to see the \emph{same} turbulence field in a characteristic advection time. Therefore, the autocorrelation becomes nonzero and the diffusivity value saturates. Figure \ref{fig:D0box} shows that increasing the computational domain size ameliorates this effect and improves data convergence to the model prediction. 



\begin{figure}
    \centering
    \includegraphics[width=0.9\textwidth]{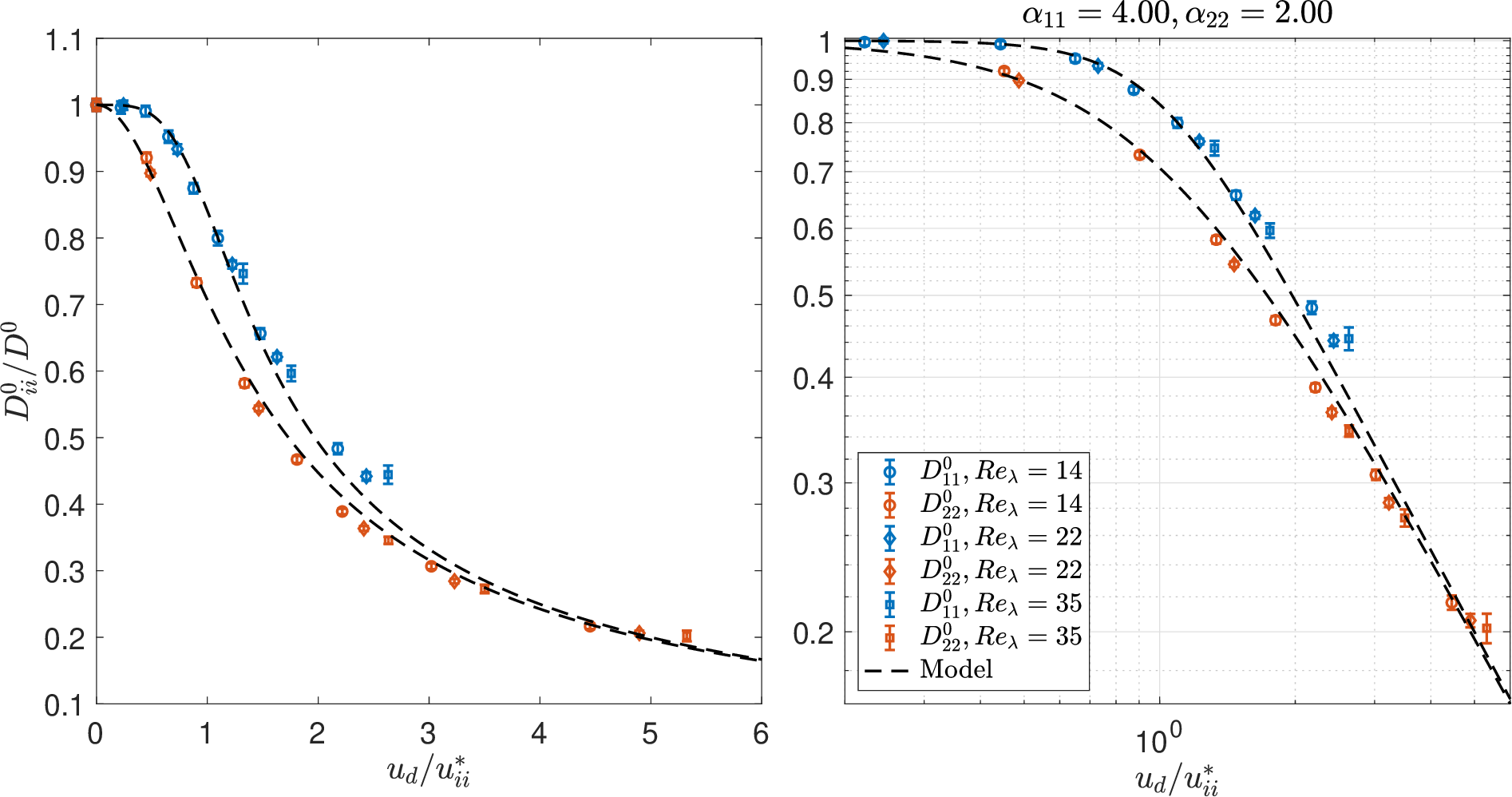}
    \caption{$D^0_{11}$ and $D^0_{22}$ for $2\pi^3$ box cases with varying $Re_\lambda$ with $\alpha_{ii}$ indicated and model predictions. Data with 95\% confidence intervals in (L) standard and (R) log-log axes.}
    \label{fig:D0Re}
\end{figure}

We now examine the effect of $Re$. In figure \ref{fig:D0Re}, $D^0_{11}$ and $D^0_{22}$ at three $Re_\lambda$ values for $u_d/u_{rms} < 5$ are shown and the data collapse when normalized by $u^*_{ii}/u_{rms}$. There is good agreement between the computational data and model predictions over the drift velocities and Reynolds numbers explored for $\alpha_{11} \approx 4$ and $\alpha_{22} \approx 2$, based on the values from figure \ref{fig:D0box}. This observed agreement is better and has narrower scatter about the model prediction than presented in \citet{squires_eaton_1991}, where $\alpha_{11}=\alpha_{22}=2$. This may be due to the long simulation time and Eulerian MFM method for quantification of $D^0_{ii}$. 

Absent drift, \citet{shirian_2022} used MFM to similarly show that scale-dependent eddy diffusivity, normalized by $u_{rms}$ and an eddy length scale, was largely invariant to $Re$. When considering $\overline{u_ic}$, this should not be surprising, as taking an average of a multi-scale field that decays rapidly in magnitude at large wavenumbers ensures that the large-scale content dominates. As $Re$ increases, figure \ref{fig:velcorr} shows us that $L_{ii}$ decreases: so $D^0_{ii}$ decreases with $Re$, \emph{ceteris paribus}. However, an increase in $Re$, properly normalized, should not greatly affect small-wavenumber quantities, which dominate the normalized measure of eddy diffusivity. 

\begin{figure}
    \centering
    \includegraphics[width=0.9\textwidth]{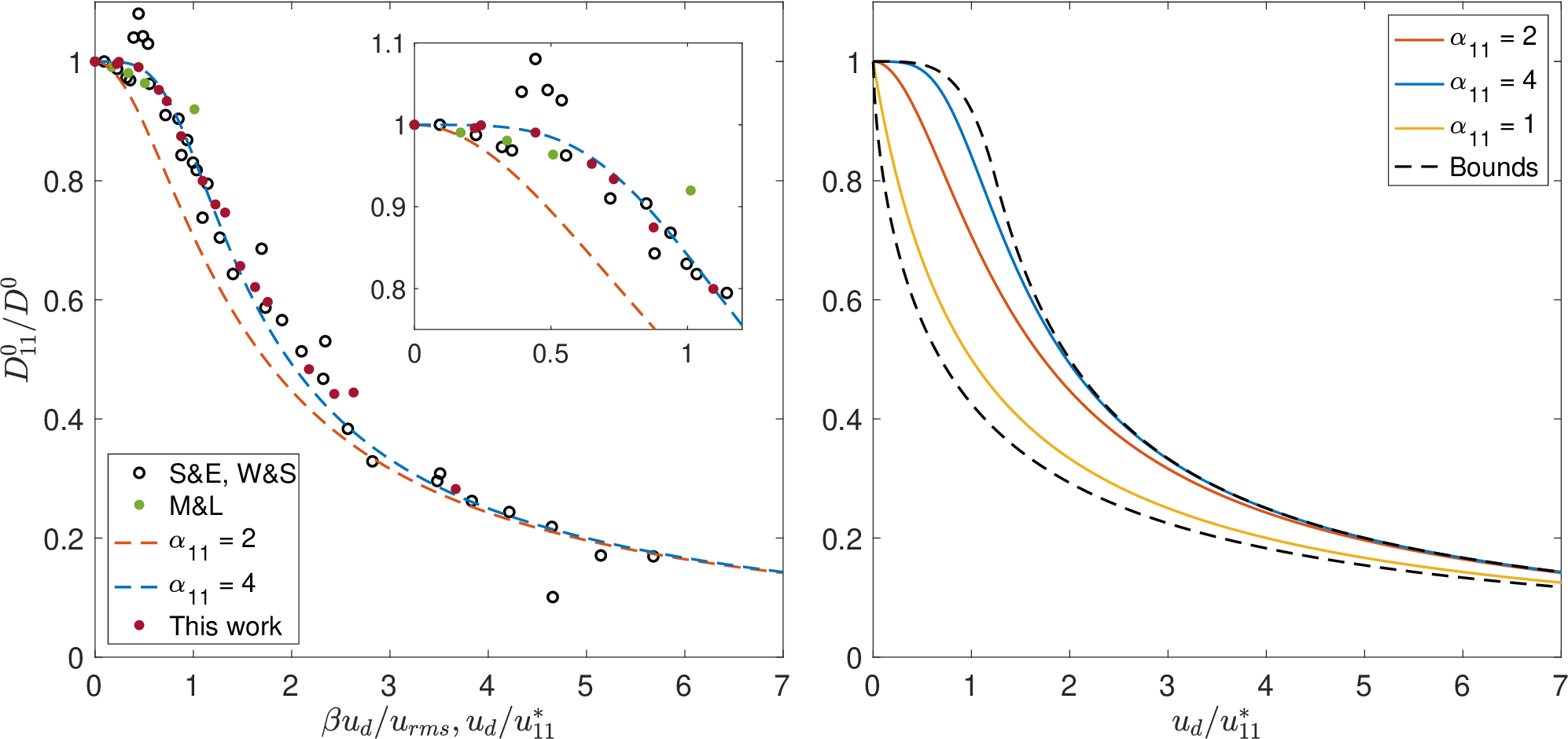}
    \caption{The present model compared to: (L) Data from this work; from \citet{wells_stock_1983} and \citet{squires_eaton_1991}; and from \citet{mazzitelli_2004} (R) Bounds at infinite $Re$ specified by \citet{yudine_1959}.}
    \label{fig:eatoncomp}
\end{figure}

The value of $\alpha_{22} = 2$ corresponds to the transverse diffusivity results of \citet{csanady_1963} and matches the overall scaling of \cite{wang_1993}, but the value of $\alpha_{11} = 4$ has not previously been reported. \citet{squires_eaton_1991} noted that their computational data and measurements of glass beads by \citet{wells_stock_1983} differed from the predictions of equation \ref{eqn:csanady}. We plot the corresponding data from \citet{squires_eaton_1991} for $D^0_{11}$ in figure \ref{fig:eatoncomp} (\textit{cf.} figure 11(b) of that work), and their version of equation \ref{eqn:csanady}, inferring $\beta = 1.1$ from their plot. If $\alpha_{11} = 4$ from this work is used, we better predict their presented data. 

\citet{squires_eaton_1991} hypothesized that discrepancies between their data and Csanady's model might be due to assumptions about the form of velocity autocorrelation and a differing values of Lagrangian and spatial measures. The eddy lengthscale control and consistent use of Eulerian correlations mean the data in figures \ref{fig:D0box} and \ref{fig:D0Re} are not affected by these considerations.  

We can plot $D^0_{11}$ markers from figure \ref{fig:D0Re} and observe that they fall along the same curve as the previous data, showing the lower scatter of MFM data about the model prediction. To build additional confidence, we can also plot computational bubble dispersion data from figure 2 of \citet{mazzitelli_2004} on the same axes. For this one-way coupled bubble data, we approximate the $D^0$ value at zero drift to compute $u^*_{11}/u_{rms} = 1.4$. The markers of figure \ref{fig:eatoncomp} come from a wide range of $Re$ values, and yet all four datasets are better described by the model with $\alpha_{11} = 4$ for the examined range of swept drift velocities.

So the value of $\alpha_{11} = 4$ is supported by data, but it also satisfies Csanady's arguments, namely that it corresponds to an autocorrelation that supports Taylor's frozen flow hypothesis. More fundamentally, \citet{yudine_1959} calculated bounds for the value of the axial eddy viscosity as a function of drift based on the Kolmogorov-Obukhov structure function description of homogeneous turbulence. For the second-order structure function required, the refined Kolmogorov hypothesis does not significantly alter the conclusions. \citep{pope_2000} These bounds are pictured in figure \ref{fig:eatoncomp} for a $u^*_{11}$ imputed from that work and it is clear that $\alpha_{11} = 4$ is close to the envelope of realizable curves, which represent the limit state of infinite $Re$ turbulence. $\alpha_{11} \approx 1$ closely describes the corresponding lower bound. As the transverse structure function differs from the axial one by only a multiplicative constant, the bounding exponents are the same for the transverse diffusivity, so $\alpha_{22} = 2$ also satisfies theoretical constraints of \citet{yudine_1959}. Since the theoretical bounds are in the infinite $Re$ limit, the model in this work should be only a weak function of $Re$, as is suggested by figures \ref{fig:D0Re} and \ref{fig:eatoncomp}. 

To use equation \ref{eqn:D0model}, $D^0$ could be written in terms of the $k$ and $\epsilon$ of the underlying turbulence; \citet{shirian_2022} attempted to establish such a correlation and showed $D^0$ scales with the large eddy size and velocity scale. As such, one could correlate $u^*_{ii} = L_{ii} u^2_{rms}/D^0$ with $u_{rms}$ such that $u^*_{11} = 2u^*_{22} = \xi u_{rms}$ in the limit of high Reynolds and P\'eclet numbers. Table \ref{tab:D0} and figure \ref{fig:eatoncomp} suggest $\xi$ is an $O(1)$ constant, but further investigations are needed to establish the convergence of $\xi$ with $Re$.

\section{Conclusions}  

We propose an algebraic model for capturing the effect of drift velocity on the turbulent dispersion of passive scalars. This model, presented in equation \ref{eqn:D0model}, captures the asymptotic limits of zero and infinite drift velocities exactly and the single free parameter of $\alpha_{ii}$ captures the effects of intermediate drift velocities. By measuring $\alpha_{11} \approx 4$ and $\alpha_{22} \approx 2$ from the data and predicting eddy diffusivity across the span of drift velocities, this work performs \emph{a priori} modelling without assuming a form for the underlying velocity structure. In so doing, we have shown a correspondence between MFM, an efficient Eulerian method for determining eddy diffusivity and other similar transport operators, and the classical Lagrangian definition.

In the model form, $D^0$ and $u^*$ capture the $Re$ dependence, and can be measured independent of drift. As the $\alpha_{ii}$ that best describes the two directions are different, the transition between the limiting asymptotic behaviours appear to occur more rapidly in the transverse directions than in the axial direction, even accounting for the differences in eddy size.

There is hope these results, which show improvements from previous models, can be applied to a wider range of situations than expected. As particle loading and Stokes number increase, the velocity field seen by a particle no longer resembles the undisturbed Eulerian field and there is two-way coupling. However, \citet{mathai_2018} found unequal axial and transverse diffusivity at volume fractions as high as \SI{5e-4}{}. Some of the particles of \citet{wells_stock_1983} had dynamics significant enough to affect the reported flow and \citet{mazzitelli_2004} simulated microbubbles with added mass, lift, and drag, but they are still acceptably described by the model in figure \ref{fig:eatoncomp}. For such cases, revisiting calculations to determine $D_{ii}^0$ and $L_{ii}$ in the presence of inertial particles might improve data collapse.


While this work's computational box setup is conventionally used for Reynolds-averaged equation closure, this model can provide subgrid-scale closures in the large-eddy simulation context, as computational cells far from a wall should represent HIT. Eddy diffusivity can be measured via an auxiliary scalar equation and $u^*$ can be constructed from filtered variables.

\backsection[Funding]{Support for this work was provided by the NSF GRFP under Grant No. 1656518, the Office of Naval Research under Grant No. N00014-22-1-2323, and the Stanford Graduate Fellowships.}

\backsection[Declaration of interests]{The authors report no conflicts of interest.}

\appendix
\section{Calculation of $L_{ii}$}  

Assuming finite box size $L_{box}$, equation \ref{eqn:lengthintegral} can be written as:

\begin{equation}  
   L_{11} = \frac{1}{2u^2_{rms}} \int^{L_{box}/2}_{-L_{box}/2} \langle u_1(x_1, x_2, x_3,t)~u_1(x_1+r,x_2,x_3,t) \rangle~dr.
\end{equation}

\begin{equation}  \label{eqn:L11definition}
   L_{11} = \frac{L_{box}}{2u^2_{rms}} \langle  u_1(x_1, x_2, x_3,t) \overline{u_1}^{x_1} \rangle = \frac{L_{box}}{2u^2_{rms}} \langle (\overline{u_1}^{x_1})^2 \rangle,
\end{equation}

\noindent where $ \overline{\bullet}^{x_i} $ denotes an average over $x_i$. $L_{22}$ and $L_{33}$ are then:

\begin{equation*}  
    L_{22} = \frac{L_{box}}{2u^2_{rms}} \langle (\overline{u_1}^{x_2})^2 \rangle,~~~~~L_{33} = \frac{L_{box}}{2u^2_{rms}} \langle (\overline{u_1}^{x_3})^2 \rangle.
\end{equation*}






\bibliographystyle{jfm}

\bibliography{bib}
\end{document}